\documentclass[prl,twocolumn,superscriptaddress,showpacs,preprintnumbers,amsmath,amssymb]{revtex4}
\usepackage{graphicx}% Include figure files
\usepackage{dcolumn}% Align table columns on decimal point
\usepackage{bm}% bold math
\usepackage{natbib}
\usepackage{amsmath}
\usepackage{amssymb}
\usepackage{here}
\begin{document}

\author{Stefano Roddaro}
\email{s.roddaro@sns.it}
\affiliation{NEST, Istituto Nanoscienze-CNR and Scuola Normale Superiore, Piazza S. Silvestro 12, I-56127 Pisa, Italy}
\author{Andrea Pescaglini}
\affiliation{{IIT}@NEST, Center for Nanotechnology Innovation, Piazza S. Silvestro 12, I-56127 Pisa, Italy}
\affiliation{NEST, Istituto Nanoscienze-CNR and Scuola Normale Superiore, Piazza S. Silvestro 12, I-56127 Pisa, Italy}
\author{Daniele Ercolani}
\affiliation{NEST, Istituto Nanoscienze-CNR and Scuola Normale Superiore, Piazza S. Silvestro 12, I-56127 Pisa, Italy}
\author{Lucia Sorba}
\affiliation{NEST, Istituto Nanoscienze-CNR and Scuola Normale Superiore, Piazza S. Silvestro 12, I-56127 Pisa, Italy}
\author{Fabio Beltram}
\affiliation{NEST, Istituto Nanoscienze-CNR and Scuola Normale Superiore, Piazza S. Silvestro 12, I-56127 Pisa, Italy}
\affiliation{{IIT}@NEST, Center for Nanotechnology Innovation, Piazza S. Silvestro 12, I-56127 Pisa, Italy}

\title{Manipulation of electron orbitals in hard-wall InAs/InP nanowire quantum dots}

\begin{abstract}
We present a novel technique for the manipulation of the energy spectrum of hard-wall InAs/InP nanowire quantum dots. By using two local gate electrodes, we induce a strong electric dipole moment on the dot and demonstrate the controlled modification of its electronic orbitals. Our approach allows us to dramatically enhance the single-particle energy spacing between the first two quantum levels in the dot and thus to increment the working temperature of our InAs/InP single-electron transistors. Our devices display a very robust modulation of the conductance even at liquid nitrogen temperature, while allowing an ultimate control of the electron filling down to the last free carrier. Potential further applications of the technique to time-resolved spin manipulation are also discussed.
\end{abstract}

\maketitle

The metal-seeded growth of semiconductor nanowires (NWs) has emerged as a flexible and promising technology for the fabrication of self-assembled nanostructures, with a potential impact on innovative device applications~\cite{LieberMRSB2003,Thelander2006}. Different materials can be easily integrated inside individual high-quality single-crystal NWs, with significantly looser lattice matching constraints with respect to alternative growth techniques~\cite{BjorkNL2002, JiangNL2007, CaroffS2008, ErcolaniN2009, Lugani2010}. NW technology thus offers a unique research and development platform for what concerns both fundamental physics~\cite{Nilsson2009, FuhrerPRL2008, RoddaroPRL2008, Doh2005, XiangNN2006, RoddaroNR2010} as well as scalable electronics~\cite{WigFET1, WigFET2, RoddaroAPL2008} and photonics~\cite{NWLED1, NWLED2} applications. NW-based single-electron devices - obtained either by using local gating or epitaxial barriers - have been so far one of the applications where this nanofabrication technology excelled, leading to device architectures where a control of electron filling down to the last free electron can be routinely obtained~\cite{FuhrerPRL2008, BjorkNL2004, BjorkPRB2005, FuhrerNL2007}. While tightly-confined single-electron systems based on NWs have been extensively demonstrated~\cite{BjorkNL2004, BjorkPRB2005, FuhrerNL2007}, their tunability has so far been limited and it has not been possible to indefinitely scale their dimensions and operating temperature because NWs with diameters below $\approx 20\,{\rm nm}$ quickly become insulating due to carrier depleted induced by quantum confinement. Various approaches to high-temperature Coulomb blockade, based for instance on ultranarrow etched Si wires~\cite{Hu2002,Shin2010}, have been demonstrated in the past. The InAs/InP NW technology, however, still offers an unmatched level of control and flexibility for what concerns the dot properties as well as its electronic filling.

\begin{figure}[h!]
\begin{center}
\includegraphics[width=0.48\textwidth]{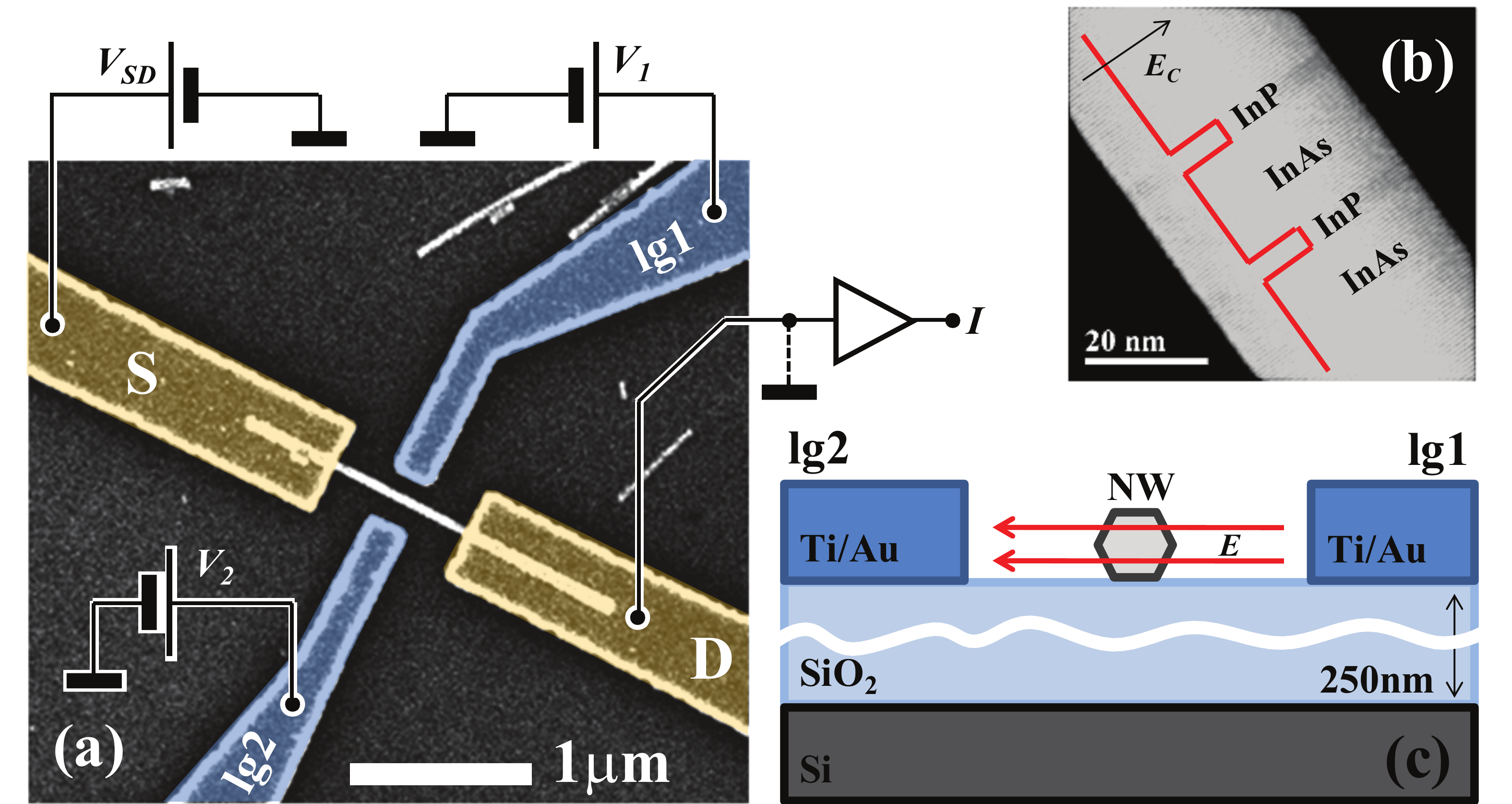}
\caption{(a) Scanning electron micrograph of one of the studied devices with a sketch of the measurement set-up in overlay. The nanowire is deposited on a SiO2/Si substrate which can act as a backgate and it is contacted by two Ti/Au electrodes (yellow). Two gate electrodes at the two sides of the nanowire (blue) allow to control the electronic filling {\em and} the dipolar field in the dot region. Gates are aligned to the position of the heterostructured dot. (b) Sanning transmission electron microscopy picture of one of the InAs/InP nanowires utilized for this research work. (c) Cross-sectional view of the device. The two lateral electrodes can be used to control the electron filling in the dot {\em and} induce a dipolar field.}
\end{center}
\end{figure}

Here we demonstrate that an electrical dipole moment can be used to easily manipulate the orbitals of an InAs/InP quantum dot and to dramatically modify its energy spectrum. In particular we show how the Coulomb gap separating the first two quantum levels in the dot can be continuously boosted up to about $75\,{\rm meV}$. Our experimental approach demonstrates how the operating temperature of a single-electron transistor is not necessarily linked to the actual dimensions of the nanostructure and that, rather, strongly-confined electron systems and high working temperatures can be obtained even in devices based on NWs of easily-manageable dimensions. Perspectives of an all-electrical manipulation of wavefunctions and energy spectrum in InAs/InP quantum dots are discussed as well.

A prototypical device is depicted in Fig.~1. InAs/InP NWs are deposited on a Si/SiO$_2$ ($250\,{\rm nm}$-thick oxide) substrate where they are contacted by aligned e-beam lithography (see Fig.~1a). Wires are fabricated using a chemical beam epitaxy process seeded by metallic nanoparticles obtained from thermal dewetting of a Au thin film~\cite{ErcolaniN2009}. Growth is performed at $420\,^o{\rm C}$ using trimethylindium (TMIn), tertiarybutylarsine (TBA), and  tributylphosphine (TBP). The TBAs and TBP are thermally cracked at around $1000\,^o{\rm C}$ upon entering the growth chamber, while the TMIn decomposes on the substrate surface. The metallorganic pressures were $0.3\,{\rm Torr}$, $1.0\,{\rm Torr}$ and $1.0\,{\rm Torr}$ for TMIn, TBAs and TBP, respectively. InAs/InP and InP/InAs interfaces where realized without any interruption by switching the group V precursors. One of the grown quantum dot structures is visible in the scanning transmission electron microscope (STEM) micrograph of Fig.~1b and contains two $5\,{\rm nm}$-thick InP barriers separated by a $20\,{\rm nm}$-long InAs island. The wires used for the present investigation have a diameter of $45\pm10\,{\rm nm}$. The position of the dot inside the NW is determined based on its average distance from the Au nanoparticle, as measured from an ensemble of wires with STEM. This leads to a typical alignment error of $\pm50\,{\rm nm}$, based on NW imaging statistics and on alignment errors during the lithographic process. Ohmic contacts are obtained by thermal evaporation of two GeAu/Au ($45/45\,{\rm nm})$ source (S) and drain (D) electrodes (yellow in Fig.~1a) located at a nominal distance of $800\,{\rm nm}$. Contacts are deposited after a chemical passivation step based on a standard ${\rm NH_4S_x}$ solution~\cite{Suyatin} and annealed at $250\,^o{\rm C}$ for $30$ seconds. Two local gate electrodes (blue, lg1 and lg2 visible in Fig.~1a and 1c) are also fabricated in correspondence to the InAs/InP quantum dot position. The two nanogates are $200\,{\rm nm}$-wide, they are placed in the middle of the gap between S and D and their relative distance is $250\,{\rm nm}$. Based on scanning electron imaging, the largely typical orientation of the NW with respect of the substrate is visibile in the cross-section view of Fig.~1c. 

\begin{figure}[h!]
\begin{center}
\includegraphics[width=0.48\textwidth]{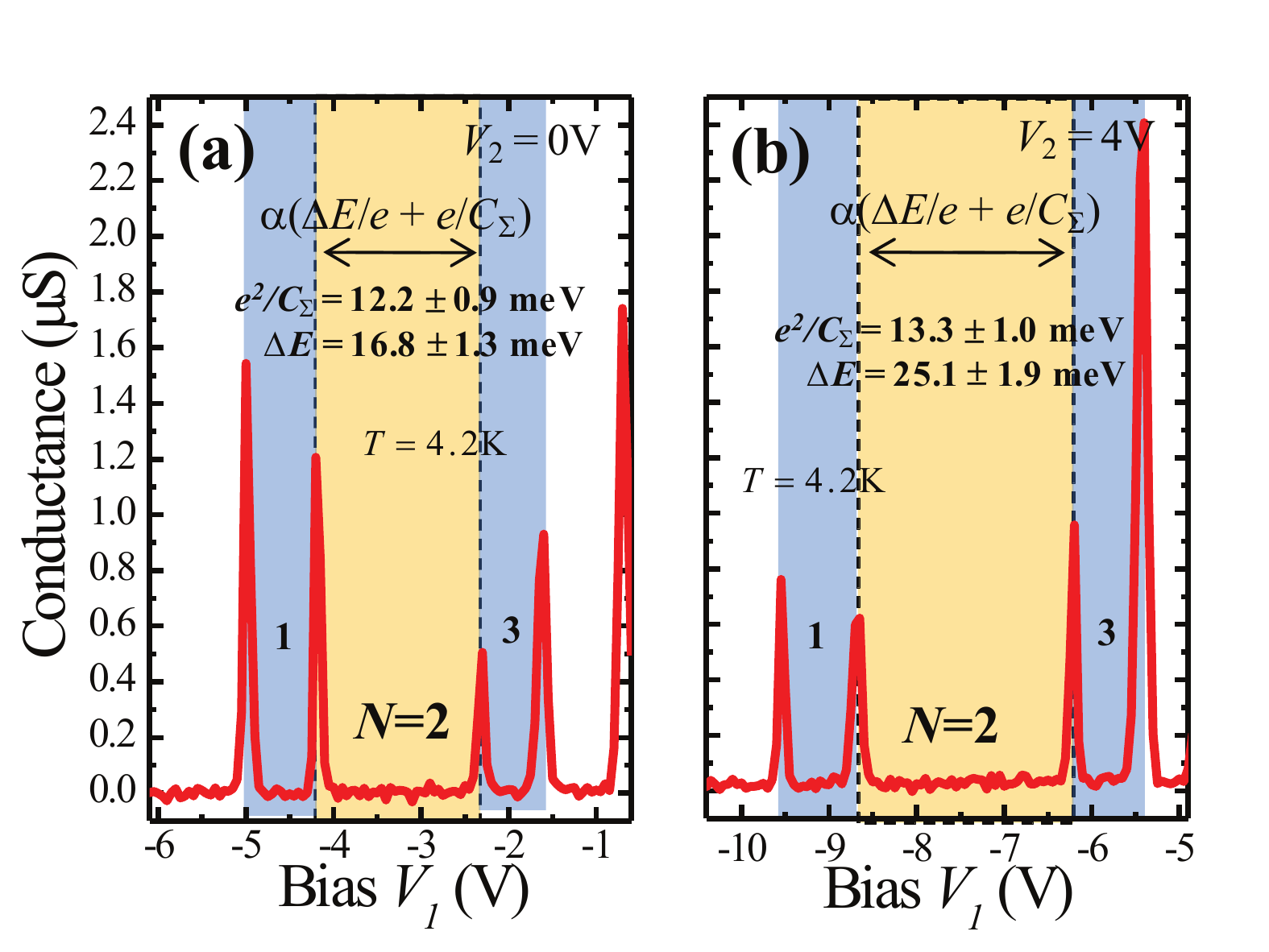}
\caption{The application of an electric dipole moment to the dot region can be used to tune the energy spectrum of confined electrons. When gate $V_2$ is set to $0\,{\rm V}$ (panel (a)) one obtains clear Coulomb blockade peaks at $T=4.2\,{\rm K}$. Transport data are interpreted in the framework of the constant interaction approximation an enhancement of the single-particle energy spacing at $N=2$ from $\Delta E=16.8\pm 1.3\,{\rm meV}$ at $V_2 = 0\,{\rm V}$ (panel (a)) up to $\Delta E=25.1\pm 1.9\,{\rm meV}$ at $V_2 = 4\,{\rm V}$ (panel (b)). The $V_1$ scan range was adjusted in order to study the same filling configurations in both panels.}
\end{center}
\end{figure}

Our device architecture is designed to allow not only the control of the electron filling in the quantum dot, but also the manipulation of its orbitals and energy spectrum. This can be achieved by applying a relative bias between lg1 and lg2 and thus by adding an electrical dipole moment to the quantum dot potential and implement a quantum-confined equivalent of the Stark effect in atoms. Given the absence of surface depletion for InAs, we expect the dot potential landscape and energy spectrum to be strongly affected by the presence of a dipole moment. This is in contrast to more standard quantum dots defined by a smooth, approximately harmonic, confinement potential, whose energy spectrum would be hardly modified by this sort of perturbation. The effect of differential biasing of lg1 and lg2 is shown in Fig.~2, comparing two Coulomb blockade scans obtained at $T=4.2\,{\rm K}$ for $V_2=0\,{\rm V}$ (Fig.~2a) and $V_2=4\,{\rm V}$ (Fig.~2b). The two scans correspond to the last $N$ free electrons in the dot~\cite{Note1} and the $V_1$ sweep range was adjusted in order to explore the same electronic fillings in both biasing configurations. In the following, our analysis will be based on the study of Coulomb gaps in the constant interaction approximation~\cite{Beenakker1991}. Gate voltage differences in the plots of Fig.~2a and 2b are converted into energies using the lever arm $\alpha_1=15.7\pm1.2\,{\rm meV/V}$ for gate lg1, which we measure by standard finite bias transport. In the left panel configuration ($V_2=0\,{\rm V}$), we estimate a charging energy $e^2/C_\Sigma = 12.2\pm0.9\,{\rm meV}$ (values extracted from the blockade regions at $N=1$ and $N=3$, blue regions), corresponding to a total capacitance $C_\Sigma = 13.1\pm1.0\,{\rm aF}$. From the Coulomb gap at $N=2$ we can thus estimate the energy distance between the first two quantum levels in the dot $\Delta E = 16.8\pm1.3\,{\rm meV}$. Both energies increase when lg2 is biased to $V_2=4\,{\rm V}$ and the lg1 scan range is adjusted accordingly: the charging energy becomes $e^2/C_\Sigma=13.3\pm1.0\,{\rm meV}$ ($C_\Sigma =  12.0\pm0.9\,{\rm aF}$) while a significantly enhanced level spacing $\Delta E=25.1\pm1.9\,{\rm meV}$ can be extracted from the data. This large change in $\Delta E$ can be understood as a consequence of the field-induced distortion of the electron orbitals in the quantum dot, as argued in details in the following.

\begin{figure}[h!]
\begin{center}
\includegraphics[width=0.48\textwidth]{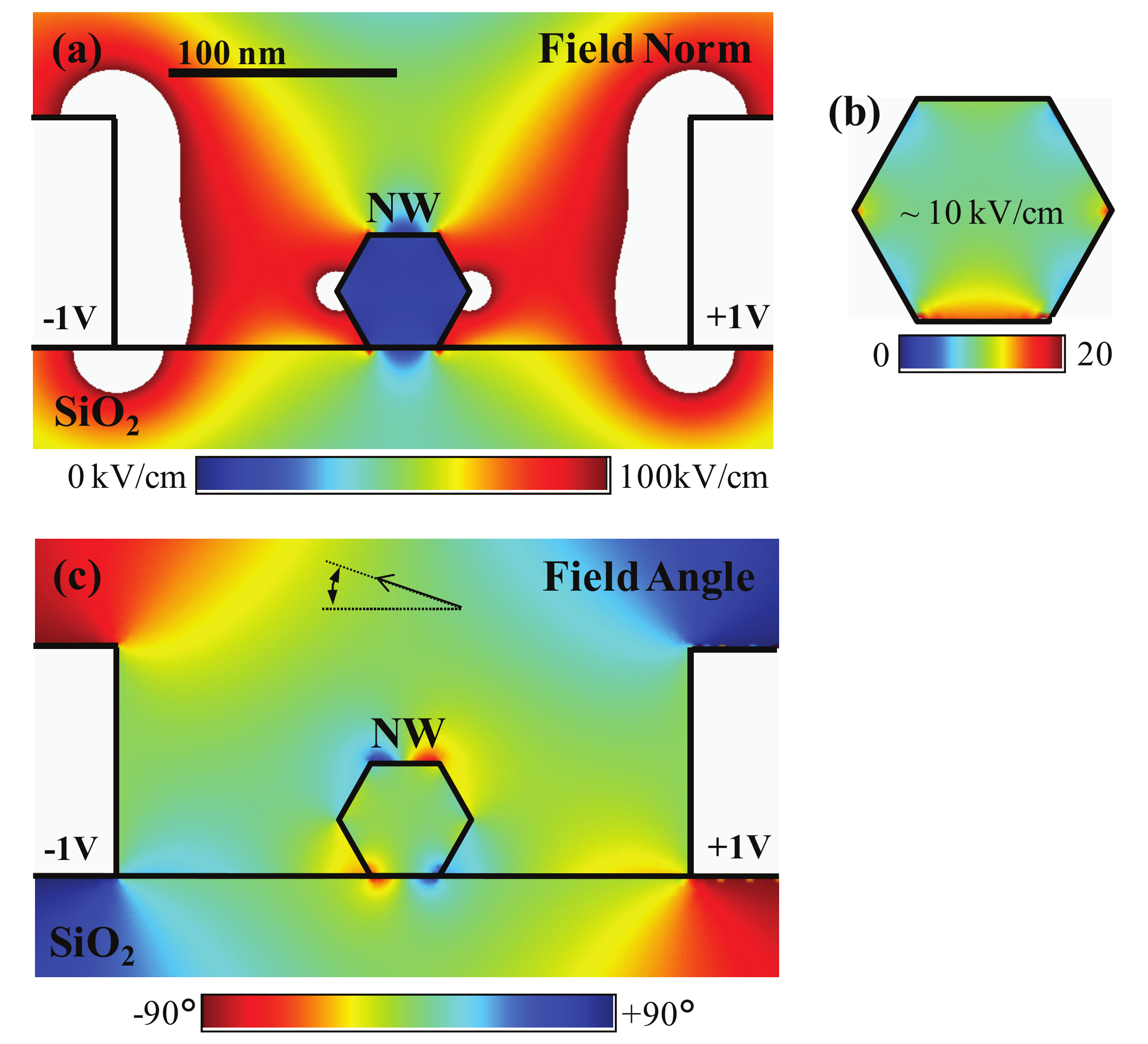}
\caption{Finite element simulation of the electrostatic field in the NW when a differential voltage $\pm1\,{\rm V}$ is applied to the two local gates. The field norm is visible in panel (a) and (b): approximately $10\,{\rm kV/cm}$ are expected inside the NW in this electrostatic configuration. Despite the complex geometry and the large dielectric mismatch between the NW and the surrounding vacuum, the induced field is expected to be, in good approximation, horizontal (panel (c)).}
\end{center}
\end{figure}

In order to better understand and analyze the effect of an independent biasing of lg1 and lg2, we have performed a finite element electrostatic simulation of our device geometry using an electrostatic solver~\cite{COMSOL}. Figure 3 illustrates the main results of the simulation, which included a $50\,{\rm nm}$-wide NW with relative dielectric constant $\epsilon_r\approx15$ and the dielectric SiO$_2$ substrate ($\epsilon_r=3.9$) and the two local gate electrodes lg1 and lg2. A differential bias $V_{1,2}=\pm1V$ is applied to the two gates. As expected from the large dielectric mismatch between the NW and the surrounding vacuum, a relatively small field is actually transferred to the NW (see Fig.~3a and 3b): even if a rough estimate based on a parallel plate capacitor approximation would lead to a field of about $80\,{\rm kV/cm}$ ($2\,{\rm V}$ over a $250\,{\rm nm}$ gap), barely $\approx10\,{\rm kV/cm}$ are expected to fall on the InAs/InP electronic island. The angle color plot of Fig.~3c also shows that the field is, in good approximation, oriented horizontally: the average absolute angle between the electric field and the horizontal axis in the NW section is found to be $\approx6.6\,{\rm ^\circ{}}$. While one should always be aware of non-ideal aspects of real devices, in particular because of surface-induced disorder potentials, finite length of the gates and possible misalignments between the gates and the dot, results of this simulation will be used as a guideline to estimate the actual field experienced by electrons in the InAs/InP quantum dot. 

\begin{figure}[h!]
\begin{center}
\includegraphics[width=0.45\textwidth]{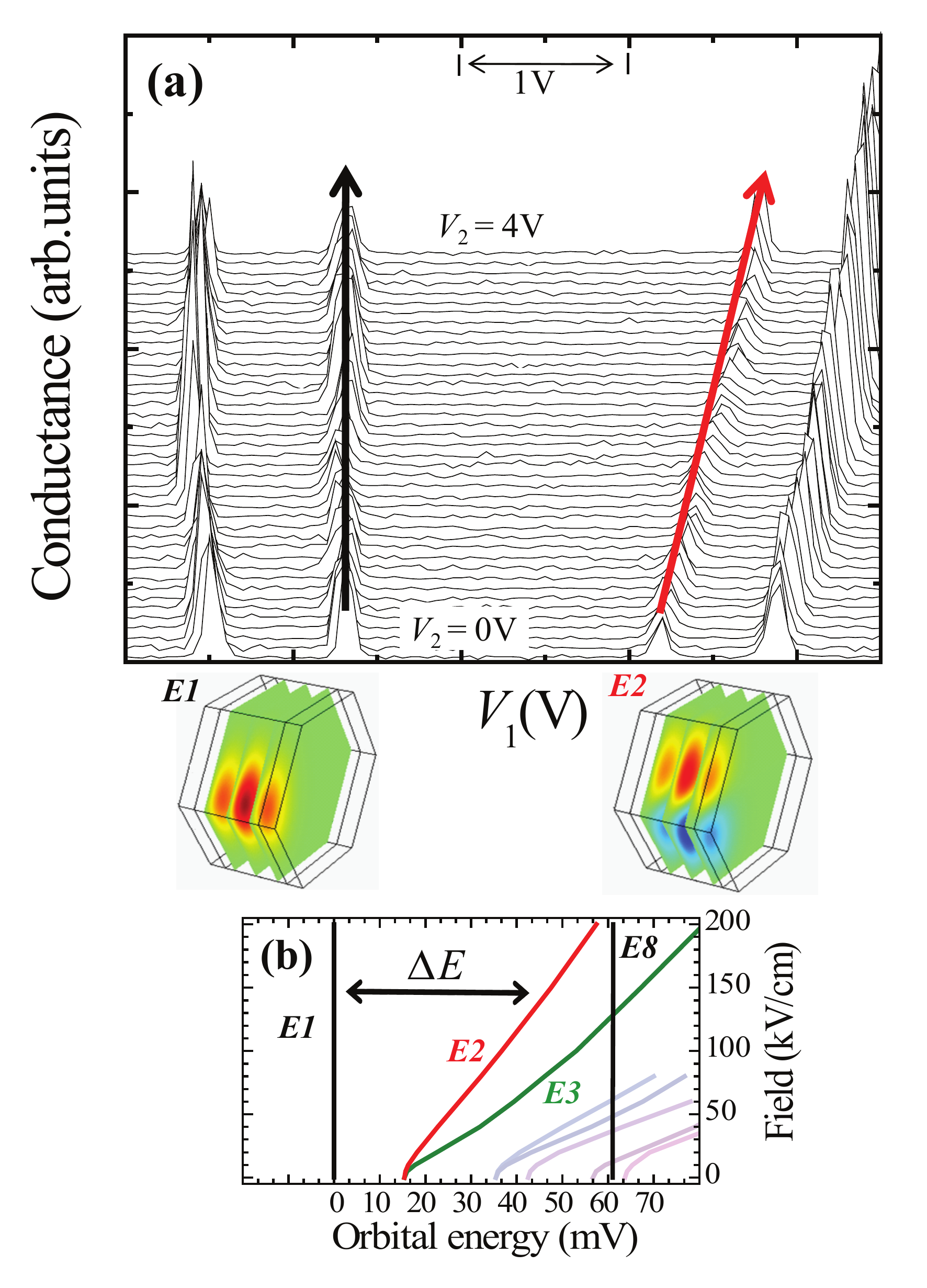}
\caption{(a) The application of a dipolar field on the quantum dot enhance the Coulomb gap at electronic filling $N=2$. (b) Evolution of the single particle spectrum of an hexagonal quantum dot as a function of a dipolar electrical field. At the bottom of panel (a), calculated wavefunctions corresponding to the first and the second dot orbitals in presence of an electric field of $100\,{\rm kV/cm}$ are reported.}
\end{center}
\end{figure}

The mechanism of level-spacing enhancement is analyzed in details in Fig.~4, where we report the full evolution of the Coulomb blockade peaks as a function of $V_2$ ($V_1$ scans are incrementally shifted for clarity, by fixing the position of the first two conductance peaks) in comparison with predictions for the energy levels of the InAs/InP quantum dot in presence of an electric dipole moment. Experimentally, the Coulomb gap increases about linearly with the bias imbalance between the two electrodes lg1 and lg2 (Fig.~4a). This is consistent with the effects of an electrical dipole on the dot energy levels, as visible in Fig.~4b. In the plot we report the single-particle energy spectrum - using the first level (E1) energy as a reference - for one electron confined inside an hexagonal InAs island with a $50\,{\rm nm}$ diameter (defined as the distance between two opposite hexagon flats) and a barrier-to-barrier distance of $16.6\,{\rm nm}$. Following results of Fig.~2, the calculation was performed~\cite{COMSOL} by adding the hard-wall dot potential to a uniform electrical field, oriented from one corner of the hexagon to the opposite one. This perturbation leads, not surprisingly, to a resolution of the intrinsic degeneracies of the hexagonal confinement (for instance, the one involving levels E2 and E3) and, in general, to an increase of the energy spacing between the dot levels. In particular a strong enhancement is predicted for the energy difference between E1 and the second level E2. This expectation can be compared with experimental data by looking at the evolution of the Coulomb gap at $N=2$. The first two Coulomb peaks indeed correspond to transport through the two spin modes of E1, while the second two involve the E2 level. The calculated spatial wavefunctions of level E1 and E2 for a field of $100\,{\rm kV/cm}$ are also reported below the corresponding peaks of Fig.~4a. The energy distance between the first two levels has already been calculated based on data at the extreme values $V_2=0\,{\rm V}$ and $V_2=4\,{\rm V}$ shown in Fig.~2 and resulting values are used here as well. The approximate transverse field at $N=2$ and $V_2=0\,{\rm V}$ can be estimated to be $\approx16\,{\rm kV/cm}$, as one can calculate starting from results of Fig.~3 and considering that $V_1=-3.25\,{\rm V}$ at the center of the $N=2$ blockade region. Such field value is ramped up to $\approx60\,{\rm kV/cm}$ for the configuration $V_2=4\,{\rm V}$ ($V_1=-7.42\,{\rm V}$ at the center of the $N=2$ blockade region). Based on calculations reported in Fig.~4b, $\Delta E\approx17\,{\rm meV}$ and $\Delta E\approx27\,{\rm meV}$ would be expected for the $V_2=0\,{\rm V}$ and $V_2=4\,{\rm V}$ configurations, respectively. The observed enhancement of $\Delta E$ is thus quite consistent with the proposed interpretation. 

It is worth remembering that intrinsic limits to the studied gap enhancement mechanism exist. As also visible from the plot of Fig.~4b, the first energy gap can in principle only be increased up to the energy required to excite the second axial mode of the dot (for the chosen geometry, E8). The energy spacing between E1 and E8 is clearly unaffected by the electric dipole moment, which shifts in the same way both states, and thus constitutes an upper bound for $\Delta E$. However, this limit can in principle be suitably loosened by designing a shorter quantum dot. Other factors that are expected to play an important role in a real device are Zener tunneling in the NW and dielectric breakdown phenomena, in general. A significant interband tunneling amplitude can surely hamper present analysis as it will invalidate the single-band approximation on which we base our interpretation in terms of conduction band quantum states. In particular, given the wurtzite InAs gap value ($\approx480\,{\rm meV}$~\cite{Wurzite}) our $45\pm10\,{\rm nm}$-wide NWs will start to accommodate a sufficient band banding to make interband tunneling possible at a field $\approx100\,{\rm kV/cm}$, even if quantum confinement effects are expected to substantially increase this threshold value. Differently, it is important to stress that in our experiment the dielectric breakdown field for bulk InAs (barely $\approx40\,{\rm kV/cm}$~\cite{Breakdown}) is not expected, nor observed, to be crucial because of the lack of actual conduction in the field direction as well as because of quantum confinement and limited lateral dimensions of the NW in general. 

\begin{figure}[h!]
\begin{center}
\includegraphics[width=0.48\textwidth]{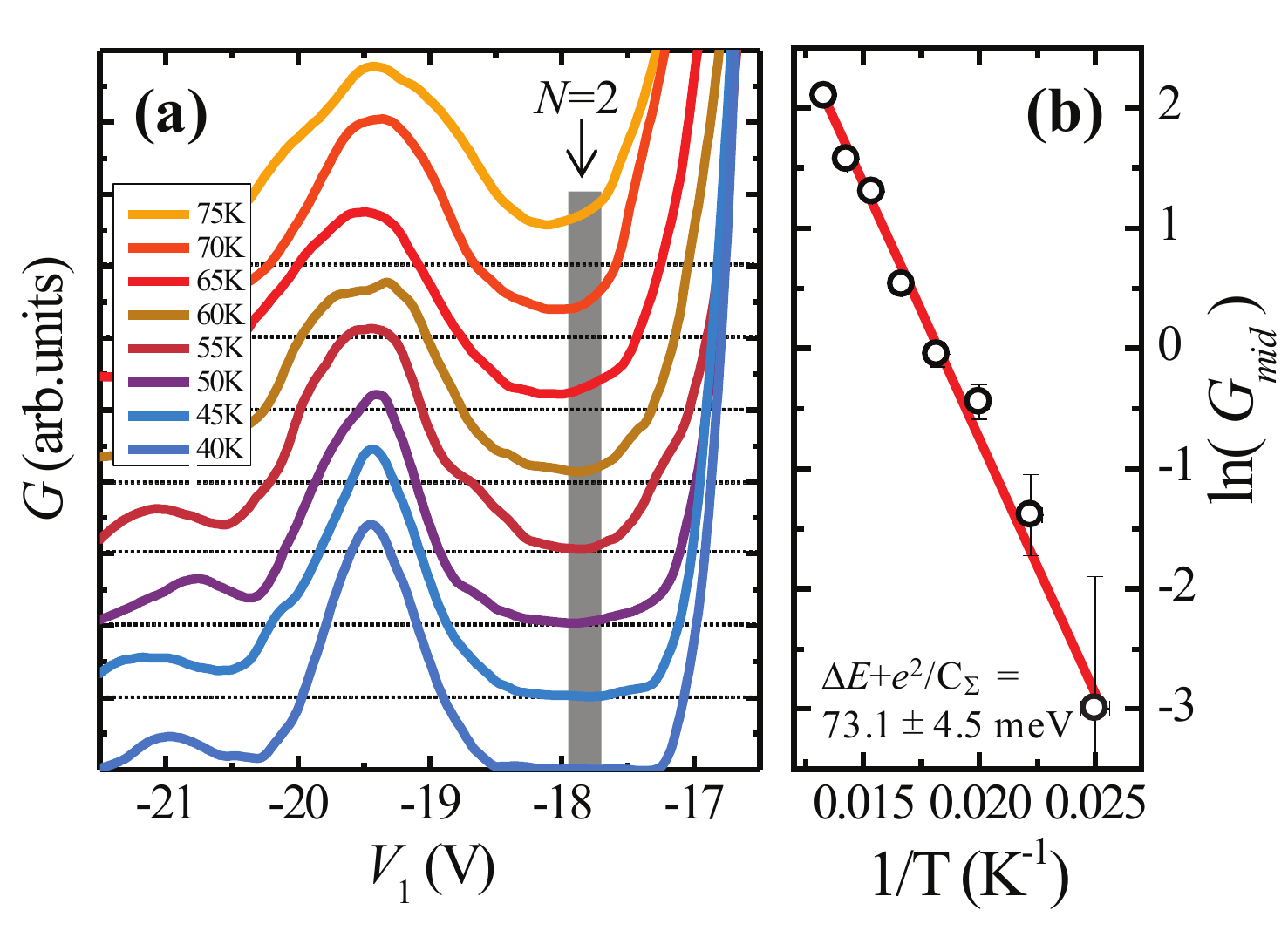}
\caption{(a) High temperature Coulomb blockade oscillations. A well-developed blockade region is observed up to a temperature $T=50\,{\rm K}$ for an electronic filling $N=2$. A very robust modulation of the conductivity is still present at liquid nitrogen temperature. (b) Arrhenius plot of the logarithm of conductance at the center of the $N=2$ region ($G_{mid}=G(-17.87\,{\rm V})$) as a function of the inverse temperature $1/T$. A Coulomb gap of $73.1\pm4.5\,{\rm meV}$ can be extracted from the data, in agreement with estimates based on finite-bias transport.}
\end{center}
\end{figure}

% Lever-arm was evaluated based on dataset 29 on the second device...

Despite the discussed limitations, a significant $\Delta E$ enhancement can be obtained using the described technique. Figure 5 shows a set of high-temperature Coulomb blockade scans obtained on a different device where a bias $V_2=12\,{\rm V}$ is applied on lg2 and $N=2$ is observed at $V_1\approx-18\,{\rm V}$. Given results of the simulation of Fig.~3, we estimate a field strength in the NW of $\approx 150\,{\rm kV/cm}$, leading to a guess expectation $\Delta E\approx50\,{\rm meV}$ (see Fig.~5b) and well-developed Coloumb blockade regions are visible for $N=2$ up to a temperature of $T=50\,{\rm K}$ (see Fig.~5a) and a very robust modulation of the conductance survives above the liquid nitrogen temperature. Finite bias spectroscopy indicate a lever arm $\alpha_1=24.3\pm3.1\,{\rm meV/V}$ and given the width of the $N=2$ region of $3.15\,{\rm V}$ (from the $G$ peak value $V_1=-19.45\,{\rm V}$ to $-16.3\,{\rm V}$, not visible in the plot) we can determine the Coulomb gap to be $76.5\pm9.8\,{\rm meV}$. This result can be cross-checked with thermally activated transport in the blockaded region $N=2$. Figure 5b reports an Arrhenius plot of the conductance $G$ at $N=2$ at the mid point between the two consecutive Coulomb blockade peaks ($G_{mid}$ at $V_1=-17.87\,{\rm V}$) as a function of the inverse of the temperature $1/T$. Based on the weak-coupling approximation~\cite{Beenakker1991}, largely valid in the studied high-temperature regime, data points are expected to follow the scaling law

\begin{equation}
G_{mid}\propto \exp\left[-\frac{e^2/C_\Sigma+\Delta E}{2k_BT}\right].
\end{equation}

Experimental points indeed fall nicely on a linear fit with $e^2/C_\Sigma+\Delta E=73.1\pm4.5\,{\rm meV}$, consistently with the estimate based on finite-bias transport.

In conclusion, we have demonstrated that the application of an electrical dipole moment to an InAs/InP nanowire quantum dot can be efficiently used to strongly modify its energy spectrum. The technique was exploited to enhance the Coulomb gap at $N=2$ up to $\approx75\,{\rm meV}$ and to significantly extend the device operating temperature. Well-developed blockade regions could be observed up to $T\approx50\,{\rm K}$ and a very robust conductance modulation survives at the liquid nitrogen temperature. More in general, we argue that the proposed technique can be useful in order to achieve a full-electrical control of the energy spectrum of InAs/InP quantum dots, to tune its energy gaps and to induce or resolve orbital degeneracies. A similar manipulation of the electronic orbitals, obtained by coupling the orbitals to a magnetic field, was demonstrated in the past on epitaxial pillar dots~\cite{SpinQD} and allowed the investigation of fascinating many-body and spin phenomena in artificial few-electron systems. Achieving a similar control by means of an electrical field can thus open a new window of opportunity for the time-resolved manipulation of few-electron spin configurations in InAs/InP quantum dot systems.

We gratefully acknowledge G. Signore for support with chemical passivation of the NWs and F. Rossi and G. Salviati for the STEM micrograph  reported in Fig.~1b. The work was in part supported by the INFM-CNR research project ``Acoustoelectrics on Self-Assembled One-Dimensional Semiconductors'' and by Monte dei Paschi di Siena through the projects ``Implementazione del laboratorio di crescita dedicato alla sintesi di nanofili a semiconduttore'' and ``Creazione di un laboratorio per lo sviluppo di nanodispositivi optoelettronici al THz''.


\begin{thebibliography}{9}
\bibitem{COMSOL} Calculations were performed using the partial differential equation solver COMSOL MultiPhysics\textsuperscript{\textregistered}.
\end{thebibliography}
\end{document}